\documentclass[]{interact}

\usepackage{epstopdf}

\usepackage[numbers,sort&compress]{natbib}
\bibpunct[, ]{[}{]}{,}{n}{,}{,}
\makeatletter
\def\NAT@def@citea{\def\@citea{\NAT@separator}}
\makeatother

\begin{document}


\title{Split-gasket approach to the integration of electrical leads into diamond anvil cells}

\author{
\name{Neha Kondedan\textsuperscript{a}\thanks{Email: neha.kondedan@fysik.su.se},
Ulrich Häussermann\textsuperscript{b}, and Andreas Rydh\textsuperscript{a} \thanks{Email: andreas.rydh@fysik.su.se}}
\affil{\textsuperscript{a}Department of Physics, Stockholm University, SE-10691 Stockholm, Sweden}
\affil{\textsuperscript{b}Department of Materials and Environmental Chemistry, Stockholm University, SE-10691 Stockholm, Sweden}
}

\maketitle

\begin{abstract}
Transport and heat capacity measurements under pressure must reconcile the limited available space and complicated geometry of a high-pressure cell with the need for multiple electrical connections. One solution for diamond anvil cells is to use customized diamonds with deposited electrical leads. Here, we instead address the problem through a split-gasket approach, intended for diamond anvil cells at moderate pressures and low temperature. A key component is the use of a substrate with lithographically defined leads, which enables connections to components such as thermometer, heater, and/or sample within the confined sample volume of the cell. The design includes an elaborate BeCu gasket sandwich with a preparation method that ensures electrical contact integrity. Using this configuration, we bring 12 leads to within 100\,$\mu$m of the center of the diamond anvil at a pressure of about 2 GPa, comparable to the pressure reached with a regular gasket, demonstrating the setup's capability for high-pressure experiments. The split-gasket approach may come at the cost of reduced maximum pressure, but brings versatility and reproducibility, and alleviates the experimental efforts of maintaining multiple electrical leads both intact and electrically isolated.
\end{abstract}


\section{Introduction}
Hydrostatic pressure serves as a tuning parameter in the study of material properties across various condensed matter disciplines \cite{khasanov2011tuning, gati2012hydrostatic, wu2011quantum, zhang2022uniaxial,ponkratz1999quasicrystals}. Pressure can induce effects similar to chemical doping, without compromising crystal quality, by modifying the lattice parameters \cite{gati2020hydrostatic,lü2014enhanced,moritomo1995pressure}. The resulting effects on orbital overlap, electronic correlation, and density of states may be profound \cite{hemley1998revealing,mao2016recent}. Consequently, pressure is an extended dimension, in addition to the use of magnetic fields, in the study of electrical resistivity, specific heat, AC susceptibility, and dielectric properties in a range of materials, particularly in superconductors and magnetic systems \cite{wang2014review,flores2020perspective,gor2018colloquium,woollam2012high}. Variations of the diamond anvil cell (DAC) \cite{jayaraman1983diamond}, a successor to Bridgman's opposed anvil cell \cite{bridgman1964resistance}, have been widely used to generate static pressures up to the 100\,GPa range and beyond \cite{dubrovinsky2000situ, yagi2020high, drozdov2015conventional, Dubrovinsky2012}. However, the maximum pressure is related to the chamber volume, typically limiting the attainable pressures for setups such as specific heat to the 1 - 20\,GPa range \cite{wilhelm2003ac, geballe2017ac, lortz2005evolution, dasenbrock2022second, umeo2016alternating, kubo2007specific}.

A major challenge in conducting high-pressure measurements lies in introducing electrical feedthroughs into the small pressure chamber containing the sample of interest, without causing a short circuit with the gasket, and maintaining stable contacts to the sample under high pressure. Various methods have been employed to achieve reliable electrical connections. For resistivity measurements, there exist setups where four wires are attached to the corners of a sample in a four-probe geometry, using either silver paste \cite{cui2003effects, colombier2010electrical,thomasson1997transport}, in-situ soldering \cite{gonzalez1986electrical, jaramillo2012four}, or mechanical contacts under applied stress \cite{torikachvili2009effect, erskine1987technique}. For some samples with low adhesion, contact pads are first deposited onto the surface, and the wires are then attached to these pads using silver epoxy \cite{patel1987electrical}.
Another way used is to deposit the leads onto the sample itself \cite{leong1992laminated}. Efforts have been made to introduce leads into the sample volume by depositing them onto the diamond anvil culet using sputtering or evaporation techniques \cite{grzybowski1984band, straaten1987electrical, hemmes1989synthesis, gao2005accurate}, or by advanced techniques such as 3D laser pantography or 2D projection lithography followed by epitaxial diamond chemical vapor deposition \cite{weir2000epitaxial}.

Beyond resistivity measurements, other setups under high pressure include anomalous Hall conductivity, which is conducted with leads deposited on an insulated gasket \cite{wang2019pressure}, ionic conductivity measured with electrodes on the diamond anvil \cite{wang2016determination}, and magnetic susceptibility measurements with specially fabricated leads on the diamond anvil \cite{jackson2003magneti}. AC calorimetry measurements require the incorporation of a heater and thermometer near the sample, necessitating at least four incoming leads. There are AC calorimetry setups that utilize insulator gaskets \cite{wilhelm2003ac,geballe2017ac,lortz2005evolution}, and insulated gaskets \cite{dasenbrock2022second}.
Some setups place the heater and thermometer outside the pressure chamber \cite{umeo2016alternating, kubo2007specific}, but this requires good thermal conduction within the cell, adding to the addenda heat capacity.

To prevent shorts between metallic gaskets and leads, non-metallic gaskets or insulated metallic gaskets are typically used. Non-metallic gaskets, such as amorphous boron \cite{lin2003amorphous}, boron composites \cite{rosa2016amorphous}, diamond powder-epoxy mixtures \cite{graf2011nonmetallic}, have been employed as inserts within metal gaskets or as standalone gaskets. Given the mechanical advantages of metallic gaskets, many experiments utilize insulator-coated metallic gaskets.
These include metal gaskets covered with an Al$_2$O$_3$ layer \cite{gonzalez1986electrical,erskine1987technique}, an Al$_2$O$_3$ and NaCl mixture \cite{straaten1987electrical}, a layer of Al$_2$O$_3$ followed by a Kapton foil \cite{hemmes1989synthesis}, an Al$_2$O$_3$ and epoxy mixture \cite{thomasson1997transport, jaramillo2012four}, mica and MgO powder \cite{reichlin1983measuring}, as well as Zylon fibers with diamond paste \cite{yomo2010moissanite}.

The primary issues with all of these setups and approaches include severing of the electric leads and their insulation by the edges of the diamond or the gasket hole, shorts to the gasket, and loss of contact with the sample under stress. Additionally, making contact with the sample can be challenging, depending on its surface characteristics and geometry. Non-hydrostatic pressure conditions may also fail the contacts during the experiment due to minor movements of the sample within the sample volume.
These issues can be addressed by introducing a robust substrate with deposited electrical leads that extends all the way from the center of the pressure chamber to the external connections on the outside. Unlike other setups, this substrate prevents direct contact between the leads and the diamond anvil, reducing the risk of damage due to the diamond edges.

\section{Experimental setup}

\begin{figure}[t!]
  \centering
\includegraphics[width=0.75\linewidth]{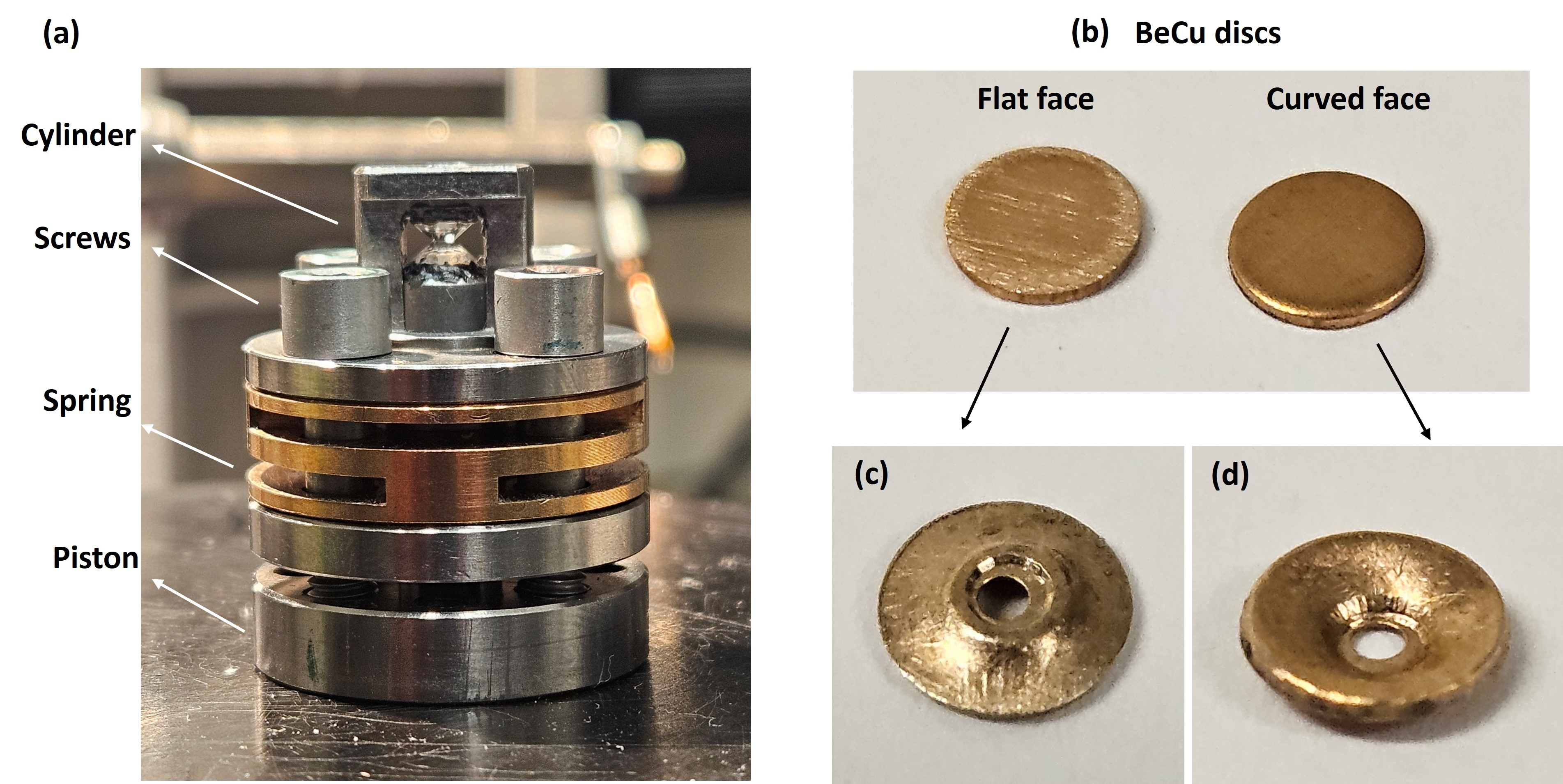}
  \caption{(a) Miniature diamond anvil cell with 28 mm height and 22 mm diameter. (b) BeCu discs of 4 mm diameter and 380 $\mu$m thickness, used as top gasket before indentation. The flat/curved asymmetry of the faces arises from the gasket punching process, used here to control the cupping direction of the gaskets. (c) Flat side of the disc after pre-indenting. (d) Curved side after pre-indenting. It is cupped as a girdle around the top anvil due to the initial curvature of the face.
  }
\label{fig:DAC and BeCu discs}
\end{figure}

Here, we use a miniature DAC (manufactured at Earth \& Planets Laboratory, Carnegie Institution for Science) weighing 39 g, with a diameter of 22 mm and a height of 28 mm, as shown in Figure~\ref{fig:DAC and BeCu discs}(a).
Its small size and low mass facilitate use with various probes and cryogenic systems. A conical aperture, with a diameter of 1 mm at the top of the cylinder, allows for optical viewing and pressure measurements. The diamond anvils are type \textsc{ii}a, 16-sided, with a culet size of 1.2 mm (supplied by Almax easyLab). The piston and cylinder design enables the application of force, which is employed by turning the four screws on top of the cell and regulated by the spring to achieve a uniform distribution of force. The top diamond is affixed to the cylinder and the bottom diamond to the piston, using Stycast 2850FT epoxy and working under optical stereoscope and viewing through the top diamond. 
The cell body is made of steel and some parts of BeCu (98\% Cu - 2\% Be). 

\section{Preparation of gaskets}
The split-gasket approach requires a thicker gasket beneath the top diamond and a thinner gasket on the bottom diamond. We use BeCu as the gasket material, as it performs well across a wide range of temperatures, particularly at low temperatures, due to its excellent mechanical properties and minimal magnetic background \cite{heller1976copper}. 
A BeCu disc with a thickness of 380 $\mu$m and a diameter of 4 mm is used as the upper gasket material, which is indented to achieve the optimal final thickness. This punched disc features two asymmetric faces — one flat and one with curved edges, as illustrated in Figure~\ref{fig:DAC and BeCu discs}(b). This asymmetry is critical for the setup. The indentation on one side leaves the opposite side flat, allowing it to face the substrate and preventing excessive deformation under pressure. For pre-indentation and hole drilling, first the BeCu disc is pre-indented to a thickness of approximately 230 $\mu$m. This process causes one face to remain flat, while the gasket cups toward the curved face, as shown in Figure~\ref{fig:DAC and BeCu discs}(c) and (d), respectively. Second, a 600 $\mu$m pilot hole is drilled at the center of the disc using electrical discharge machining.
The disc is then indented further, causing additional deformation and a reduction in the hole size. This step is essential to prevent the pressure chamber from shrinking during the experiment. The additional indentation reduces the central thickness to approximately 130 $\mu$m and decreases the hole size to around 400 $\mu$m. The hole is then redrilled to restore its original size. The surface roughness of the disc is measured by averaging the profile deviations from a scan conducted using a Stylus Profilometer KLA Tencor P7. After machining, the roughness is approximately 0.5 $\mu$m, indicating a relatively smooth surface.

The lower gasket is a 50 $\mu$m thick BeCu foil. This gasket is typically cut to match the size of the electrical leads substrate, primarily serving as support for the substrate. Due to its small thickness compared to the top gasket, the lower gasket exhibits minimal deformation under applied force \cite{timofeev2003measuring}.
Therefore, it does not require pre-indentation. The surface roughness of the foil is about 0.3 $\mu$m. Both gaskets are uninsulated.

\section{Integrating leads into the cell}
\begin{figure}[t!]
  \centering
\includegraphics[width=0.9\linewidth]{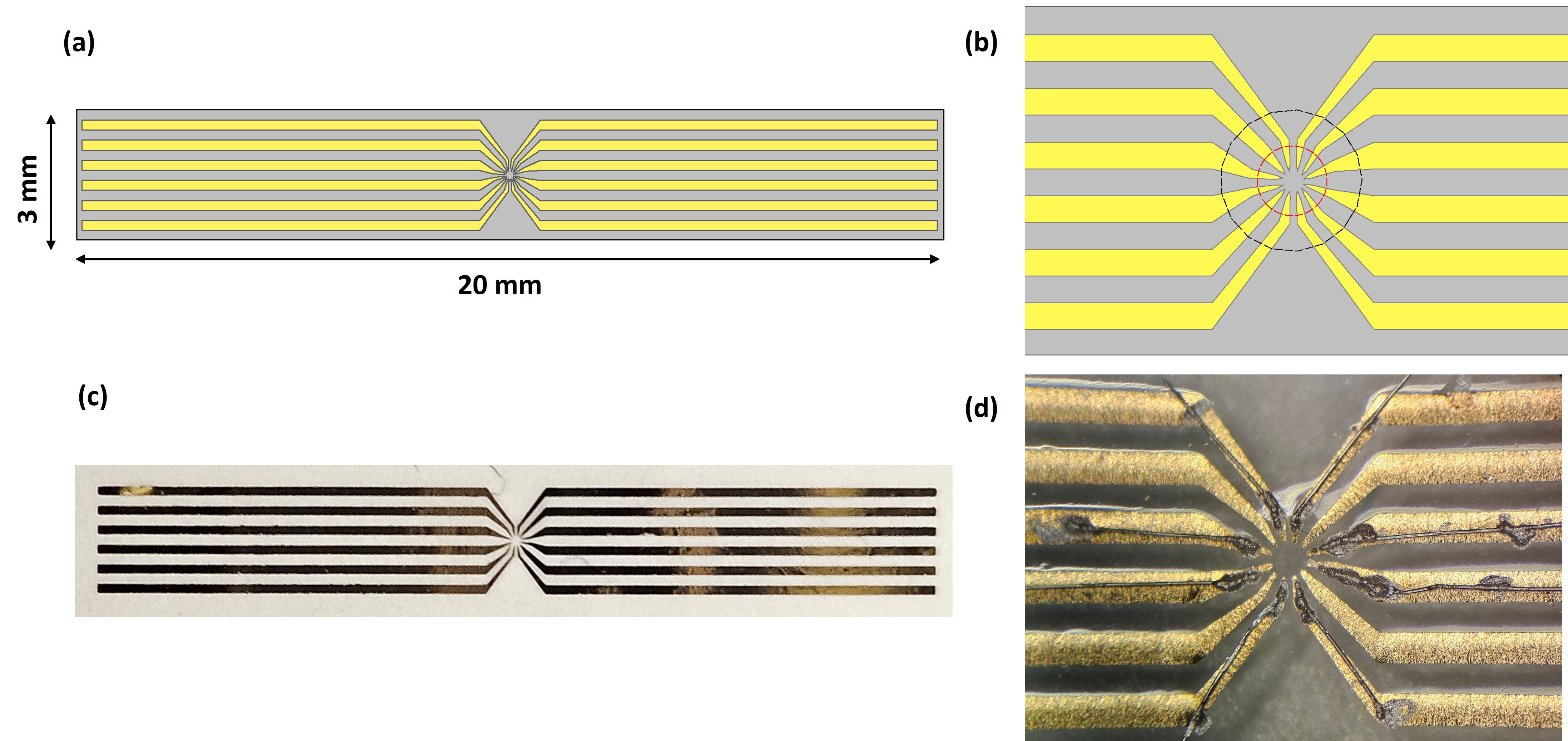}
  \caption{(a) Layout of the substrate with metallic leads on top. (b) Magnified view of the center of the substrate where the leads converge to a circle, on top of which a sample or wafer chip can be attached. The locations of the diamond culet edges (black line) as well as the gasket hole (red line) are indicated by dashed lines. (c) PEEK substrate with deposited Cr+Au leads. (d) Central area with small pieces of Al wires attached to the leads, to reinforce the electrical contacts where most part of the deformation takes place under pressure.
  }
\label{fig:substrate}
\end{figure}

The electrical leads are integrated through a thin, electrically insulating substrate with lithographically patterned leads. The leads' design allows multiple leads to be positioned close to the sample inside the sample volume. The layout of the substrate with leads is shown in Figure~\ref{fig:substrate}(a). With a length of 20 mm, the substrate extends outside the chamber to make external connections.
The design incorporates 12 leads arranged in a circular pattern at the center, with a diameter of 190 $\mu$m, as illustrated in Figure~\ref{fig:substrate}(b). Each lead has a width of 230 $\mu$m, gradually narrowing to 10 $\mu$m at the center of the circle. These leads are fabricated using nanofabrication techniques.

A non-metallic substrate material, PEEK (Polyether ether ketone), was selected for its excellent chemical, mechanical, and thermal resistance properties under high-pressure conditions. This thermoplastic polymer, with a thickness of 25 $\mu$m, is a smooth and transparent film compatible with high-vacuum environment and lithography chemistry. The electric leads are fabricated using a combination of photolithography, O$_2$ plasma ashing, e-beam evaporation, and lift-off techniques. A layer of 30 nm chromium, followed by 50 nm of gold is deposited as the metal leads. Additional thickness can be added, but care should be taken to avoid excessive heating of the PEEK. The resistance of the resulting leads is approximately 6 $\Omega$. Around 20 substrate samples of the required size can be fabricated simultaneously with a 4$^{\prime\prime}$ wafer system. Figure~\ref{fig:substrate}(c) shows a PEEK substrate with the leads after deposition.
Small strips of aluminum wires with a diameter of  25 $\mu$m are attached to these leads, strengthening the points where the diamond edges meet as well as at the edge of the drilled gasket, to prevent any rupture of the leads under stress, see Figure~\ref{fig:substrate}(d).

\section{Sample mounting}
For resistivity measurements, the sample can be directly attached to the center of the substrate, with electrical contacts for a four-probe geometry or multiterminal setups made to the sample using a two-component silver epoxy (EPO-TEK\textsuperscript{®} H20E), which is cured at 125$^\circ$C for 10 to 15 minutes. 
Alternatively, a silicon chip that is 30 $\mu$m thick and 300 $\mu$m in diameter, with deposited leads designed for resistivity measurements, shown in Figure~\ref{fig:pressurization}(a), can be attached to the center of the substrate using crystal bond. To avoid electrical shorts, the edges of the chip need to be protected by the crystal bond. The electric leads on the substrate can then be connected to the leads on the chip using silver epoxy. The sample is finally mounted on top of the chip, allowing connections to be made to the sample.

Specific heat measurements require similar preparations. The cell fits a calorimeter chip of a size similar to the silicon chip (30 $\mu$m thick and 300 $\mu$m in diameter), containing thermometer and heaters. The development of such a calorimeter chip is reported in \cite{NehaHighPCalorimeter}. The connections are made with silver epoxy as illustrated in Figure~\ref{fig:pressurization}(b). The sample is secured to the calorimeter surface using a thin layer of Apiezon-N grease or melted crystal bond.

\section{Application of pressure}
\begin{figure}[t!]
  \centering
\includegraphics[width=1\linewidth]{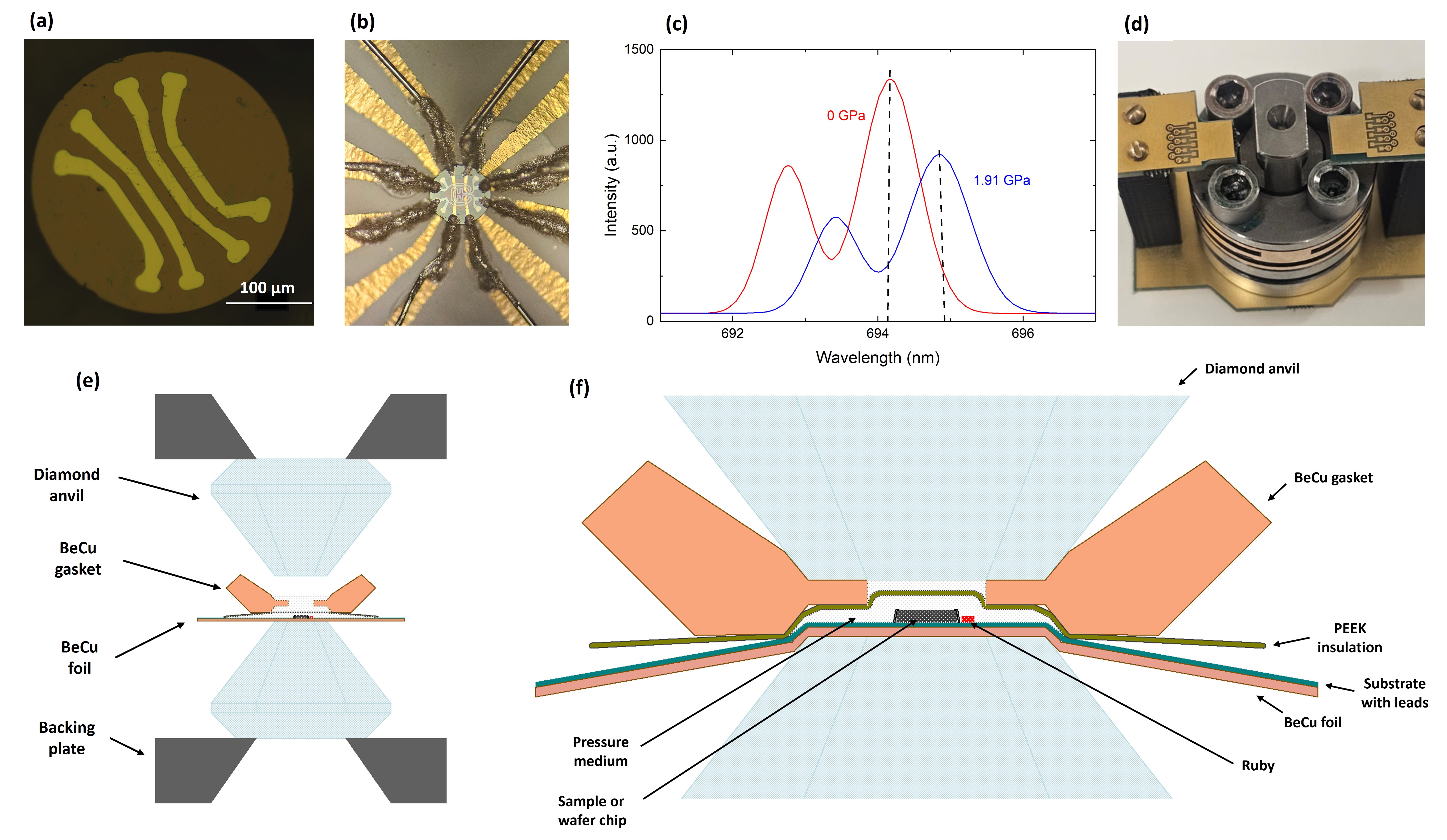}
  \caption{(a) Microscopic image of a Si chip with metal leads for transport measurements, matching the dimensions of the pressure chamber. (b) Miniaturized calorimeter attached to the center of the substrate and connected to the leads using silver epoxy. (c) Ruby fluorescence spectra showing the wavelength shift with pressure, going from 694.16\,nm at 0\, GPa to 694.85\,nm at 1.91\,GPa. (d) Bonding pads for the external electrical attachment of the substrate outside the DAC, allowing the lead ends to connect via wire bonding. (e) Schematic showing the gasket-sandwich before applying pressure. (f) Diagram showing the deformation of the gasket-sandwich components after the application of pressure.
  }
\label{fig:pressurization}
\end{figure}

The assembly of the pressure cell components is illustrated in the schematic of Figure~\ref{fig:pressurization}(e). The electrical leads substrate is placed onto the lower gasket of a similar size, with the sample space positioned at the center of the culet. Some ruby crystals are placed close to the sample to measure the pressure inside the chamber using the ruby fluorescence spectroscopy technique \cite{chijioke2005ruby}. The sample is then covered by a droplet of liquid pressure medium, which is used to maintain hydrostatic conditions. The liquid medium causes less motion and damage to single crystal samples during the loading, compared to a solid pressure medium.
With this setup, we have used Apiezon-N grease and silicone oil, but glycerol or other liquid pressure media should also work.

A 25 $\mu$m thick layer of PEEK is placed on top of the leads and sample assembly as an insulation layer to prevent shorts between the upper gasket and the substrate leads. This method of incorporating a separate insulation layer ensures that any plastic deformation in the upper gasket does not directly impact the components beneath it. The transparent PEEK also allows for easier alignment of the upper gasket.

For the alignment of the indented upper gasket with the sample space, some melted crystal bond is first applied to the flat side of the gasket prior to attaching it to the PEEK. This effectively seals any gaps between the gasket and the PEEK, preventing subsequent leakage of pressure medium contained within the gasket hole. When filling the medium, it is necessary to ensure that there are no bubbles, as they can cause volume collapse of the chamber. Once the medium is filled, the properly aligned assembly can be closed with the top diamond anvil, while checking the alignment through the optical window on top of the pressure cell.

Figure~\ref{fig:pressurization}(f) shows a diagram of the gasket assembly after pressurization. The PEEK insulation effectively acts as part of the pressure medium as long as the chamber is filled and uniformly pressurized. The maximum pressure of about 2 GPa for the culet size used can be achieved without compromising the contacts. Hydrostatic conditions are indicated by the non-broadening of the ruby fluorescence spectral lines, shown in Fig.~\ref{fig:pressurization}(c). The external leads are subsequently connected to the bonding pads shown in Figure~\ref{fig:pressurization}(d).

\section{Discussion}
The described sandwich approach presents a method of pressurizing samples while incorporating electric leads. While other methods may likely reach higher pressures, the split gasket enables various measurements without the need to change most of its components. Swapping the transport or calorimeter chip, replacing the sample, and making new connections with silver epoxy are sufficient to prepare for a new measurement.
The upper and lower gaskets, along with the substrate containing the leads, are reusable until they either break or undergo significant deformation. Additionally, this setup accommodates different culet sizes; the same substrate can be utilized with minor modifications to the dimensions of the upper gasket hole. The substrate can easily accommodate larger wafer chips and/or samples.

The process of fabricating the substrate with leads is quite simple, making it easy to alter the lead patterns as needed. With multiple leads available within the pressure chamber, some leads can be reconfigured into a heater, allowing for local Joule heating of the sample. This prevents undesired pressure changes due to temperature changes of the DAC arising from thermal expansion effects, if working at low temperature.

\section{Conclusions}
In conclusion, we have developed a high-pressure setup that can be used for various transport and calorimetry measurements requiring several electrical leads (12 demonstrated here). The key components of this approach include a split gasket and an insulating substrate with integrated leads. We find that the split-gasket solution is not limiting the pressure range of the cell that we use, reaching about 2 GPa with a 1.2 mm culet size without damaging the leads. The ease of modifying and reproducing the pressure cell components makes this setup versatile for modest pressure experiments. There is potential to reach even higher pressures by modifying the culet size and sample volume. Whether the split gasket design will limit the ultimate pressure in such a case remains to be investigated.

\section*{Acknowledgements}
Support from the Knut and Alice Wallenberg Foundation
under Grant No.\,KAW 2018.0019 and
the Swedish Research Council, Grant No.\,2021-04360, are acknowledged. We thank P.\,Lazor for fruitful discussions and H.\,Breton for introduction to the pressure cell operation.

\end{document}